\documentclass{article}
\usepackage{amsmath,amsthm,amssymb,authblk}

\newtheorem{theorem}{Theorem}[section]
\newtheorem{corollary}[theorem]{Corollary}

\newtheorem{lemma}[theorem]{Lemma}
\newtheorem{definition}[theorem]{Definition}
\newtheorem{example}[theorem]{Example}

\newtheorem{remark}[theorem]{Remark}
\numberwithin{equation}{section}
\def\pf{{\bf Proof.}~ }
\def\1.5{$1\frac{1}{2}$}

\begin{document}

\title{$\mathbb{Z}_2\mathbb{Z}_4$-Additive Cyclic Codes Are Asymptotically Good}
\author{Yun Fan,\quad Hualu Liu\\
\small School of Mathematics and Statistics\\
\small Central China Normal University, Wuhan 430079, China}
\date{}
\maketitle

\insert\footins{\footnotesize{\it Email address}:
yfan@mail.ccnu.edu.cn (Yun Fan). hwlulu@aliyun.com (Hualu Liu).}

\begin{abstract}

We construct a class of $\mathbb{Z}_2\mathbb{Z}_4$-additive cyclic codes
generated by pairs of polynomials, study their algebraic structures,
and obtain the generator matrix of any code in the class.
Using a probabilistic method,
we prove that, for any positive real number $\delta<1/3$ such that
the entropy at $3\delta/2$ is less than $1/2$,
the probability that the relative minimal distance
of a random code in the class is greater than $\delta$ is almost $1$;
and  the probability that the rate of the random code equals to $1/3$
is also almost $1$. As an obvious consequence,
the $\mathbb{Z}_2\mathbb{Z}_4$-additive cyclic codes
are asymptotically good.

\medskip
{\bf MSC classes:}~  94B15, 94B25, 94B65.

\end{abstract}

\section{Introduction}

A class of codes is said to be asymptotically good
if there is a sequence $C_1,C_2,\cdots$ of codes in the class with length
$n_i$ of $C_i$ going to infinity such that
 both the rate of $C_i$ and the relative minimum distance of
$C_i$ are positively bounded from below.
By a Varshamov's probabilistic argument \cite{V}, linear codes
over a finite field are asymptotically good.

Cyclic codes are an important class of codes and studied extensively.
It is a long-standing open question:
are the cyclic codes asymptotically good?
The question was first asked
by Assmus, Mattson, and Turyn in 1966 (see \cite{AMT}).
Many related researches are unfolded by
revolving around the question, for example, studies on asymptotic properties
of quasi-cyclic codes, group codes, and so on. A classical result is that
the quasi-cyclic codes of index $2$ are asymptotically good (see \cite{CPW}, \cite{C}, \cite{K}).
Note that quasi-cyclic codes of index $1$ are just cyclic codes.
Recently, \cite{FL16} introduced quasi-cyclic codes of fractional index,
and proved that the quasi-cyclic codes of fractional index
between $1$ and $2$ are asymptotically good.

In 1994, Hammons et al. discovered that some
good non-linear binary codes can be viewed as the Gray map
images of some linear codes over $\mathbb{Z}_4$ (see \cite{HKCSS}).
From then on, the study of codes over
$\mathbb{Z}_4$ and other finite rings has been developing.
In particular, linear and cyclic codes over $\mathbb{Z}_4$
have been studied 
 intensively (\cite{AS}, \cite{AGO}, \cite{AO}, \cite{PQ}).

In 1973, Delsarte \cite{D5}) defined additive codes
as subgroups of the underlying abelian groups in translation association schemes.
For the binary Hamming scheme, i.e., when the underlying abelian group
is of order $2^n$, only the abelian groups of the form
$\mathbb{Z}_2^\alpha\times\mathbb{Z}_4^\beta$
with $\alpha+2\beta=n$ appear.
The subgroups $C$ of $\mathbb{Z}_2^\alpha\times\mathbb{Z}_4^\beta$
are called $\mathbb{Z}_2\mathbb{Z}_4$-additive
codes (\cite{BBDF}, \cite{BFPRV}, \cite{RRR}).
If $\beta=0$, then $\mathbb{Z}_2\mathbb{Z}_4$-additive
codes are binary linear codes.
If $\alpha=0$, then $\mathbb{Z}_2\mathbb{Z}_4$-additive
codes are the quaternary linear codes over $\mathbb{Z}_4$.  So,
this class of codes contains all binary and quaternary linear codes as a subclass.
Moreover, this class of codes has shown at its early stage
some promising applications. For example,
perfect $\mathbb{Z}_2\mathbb{Z}_4$-additive codes
have been utilized in the subject of steganography (\cite{RRR}).
Therefore, 
$\mathbb{Z}_2\mathbb{Z}_4$-additive codes are an important class of codes.
In recent years, the structure and properties of $\mathbb{Z}_2\mathbb{Z}_4$-additive codes have been intensely studied (see \cite{BFPRV}).
In \cite{ASA}, Abualrub, Siap and Aydin first introduced $\mathbb{Z}_2\mathbb{Z}_4$-additive cyclic codes.
\cite{BFT} studied the generator matrices and dual codes of $\mathbb{Z}_2\mathbb{Z}_4$-additive cyclic codes.
Borges, Dougherty, Fern\'andez-C\'ordoba and Ten-Valls study the rank and the kernel of $\mathbb{Z}_2\mathbb{Z}_4$-additive cyclic codes in \cite{BDFT}.

An interesting question we are concerned with comes up:
are the  $\mathbb{Z}_2\mathbb{Z}_4$-additive cyclic codes
asymptotically good?

In this paper we answer the question positively.
A key innovation to solve the question is that: within 
$\mathbb{Z}_2\mathbb{Z}_4$-additive cyclic codes
we find a class of codes such that
the codes in the class possess good algebraic structures,  in fact, 
they are algebraically isomorphic to the binary quasi-cyclic codes 
of index two generated by pairs of polynomials; and,
though the algebraic isomorphism dos not preserve 
the rates and the relative minimum distances of the codes,  
we have a technical probabilistic method
to determine asymptotically the rates and the relative minimum distances  
of the codes in the class.
 

In Section 2, we state some preliminaries on
$\mathbb{Z}_2\mathbb{Z}_4$-additive codes and
$\mathbb{Z}_2\mathbb{Z}_4$-additive cyclic codes.
The rate $R(C)$ and the relative minimum distance
$\Delta(C)$ of a $\mathbb{Z}_2\mathbb{Z}_4$-additive code $C$
are also defined in this section.

In Section 3, we construct the class of the 
$\mathbb{Z}_2\mathbb{Z}_4$-additive cyclic codes $C_{a,b}$ (defined below)
which will play an important role as mentioned above,
By $R_m=\mathbb{Z}_2[X]/\langle X^{m}-1\rangle$
we denote the residue ring of the polynomial ring $\mathbb{Z}_2[X]$
over $\mathbb{Z}_2$ modulo the ideal $\langle X^{m}-1\rangle$
generated by $X^{m}-1$.
And $R'_{m}=\mathbb{Z}_4[X]/\langle X^{m}-1\rangle$ similarly.
We define the {\em $\mathbb{Z}_2\mathbb{Z}_4$-additive cyclic codes
with cycle length $m$} to be the $R'_m$-submodules of $ R_m\times R'_{m}$.
With $\mathbb{Z}_2\mathbb{Z}_4$ as an alphabet, 
such codes has many properties similar to the classical cyclic codes, 
e.g., they can be implemented by feed-back shift registers. 
For our purpose, we are interested in 
the $R'_m$-submodules of $ R_m\times 2R'_{m}$
where $2R'_{m}$ is the ideal of $R'_m$ generated by $2$.
Furthermore, to ignore the trivial case, we consider the ideal
$J_m=\langle X-1\rangle_{R_m}$ of $R_m$ generated by $X-1$,
and the ideal $2J'_m=\langle 2(X-1)\rangle_{R'_m}$
of $R'_m$ generated by $2(X-1)$. Then, for any polynomials $a(X),b(X)\in J_m$,
we have an $R'_m$-submodule $C_{a,b}\subseteq J_m\times 2J'_m$
generated by $(a(X),2b(X))$ in a natural way: 
$C_{a,b}=\big\{\big(c(X)a(X),2c(X)b(X)\big)\,\big|\, c(X)\in J_m\big\}$.
By investigating  the two circulant matrices associated
with the two polynomials $a(X)$ and $b(X)$,
we get the generator matrix of $C_{a,b}$ from the
polynomial pair $a(X),b(X)$. And we obtain necessary information
about the number of the ideals of $J_m$ and their dimensions 
for later using to estimate the relative minimum
weight $\Delta(C_{a,b})$ and the rate $R(C_{a,b})$.

In Section 4, we view $J_{m}\times 2J'_m$ as a probability space with
equal probability distribution, so that the
$\mathbb{Z}_2\mathbb{Z}_4$-additive cyclic codes $C_{a,b}$
with cycle length $m$ for $a(X),b(X)\in J_{m}$
is a random code, hence the relative minimum distance $\Delta(C_{a,b})$ 
and the rate $R(C_{a,b})$ are random variables. 
We study the asymptotic properties with $m\to\infty$
of the two random variables. 
Given a real number $\delta$ such that $0<\delta<1/3$ 
and the binary entropy $H(3\delta/2)<1/2$
(it is about that $0<\delta\le 0.0733$).
For each $c(X)\in J_m$, the word $\big(c(X)a(X),2c(X)b(X)\big)\in C_{a,b}$
is a random codeword which gives a Bernoulli variable $Y_c$: 
$Y_c=1$ if the random codeword is a non-zero one with relative weight $ \le\delta$, 
and $Y_c=0$ otherwise. 
Applying \cite[Theorem 3.3]{FL} which estimates the number
of the codewords in a code whose relative weights are bounded from above, 
we calculate the expectation of $Y_c$;  and then we prove that 
the probability that $\Delta(C_{a,b})>\delta$ is convergent to $1$.
Next, using an asymptotic property of the dimensions 
of irreducible modules in $J_m$, we prove that, in an asymptotic sense,
the probability that $R(C_{a,b})=\frac{1}{3}-\frac{1}{3m}$ 
is also convergent to $1$. 
Finally, it is an obvious consequence that
$\mathbb{Z}_2\mathbb{Z}_4$-additive cyclic codes are asymptotically good.

\section{Preliminaries}
First of all, we sketch briefly
{\em $\mathbb{Z}_2\mathbb{Z}_4$-additive codes $C$}.
Any subgroup $C$ of $\mathbb{Z}_2^{\alpha}\times\mathbb{Z}_4^{\beta}$
is called a $\mathbb{Z}_2\mathbb{Z}_4$-additive code.
It is obvious that $\mathbb{Z}_2^{\alpha}\times\mathbb{Z}_4^{\beta}$
is a $\mathbb{Z}_4$-module, and its subgroups are $\mathbb{Z}_4$-submodules.

As usual, the weight of
${\bf a}=(a,a')\in\mathbb{Z}_2^{\alpha}\times\mathbb{Z}_4^{\beta}$
is defined by
\begin{equation}\label{weight}
 {\rm wt}({\bf a})={\rm wt}(a,a')={\rm wt}_H(a)+{\rm wt}_L(a'),
\end{equation}
where ${\rm wt}_H(a)$ is the Hamming weight of $a$
and ${\rm wt}_L(a')$ is the Lee weight of~$a'$.
The minimum weight of the $\mathbb{Z}_2\mathbb{Z}_4$-additive code $C$
is defined as
${\rm wt}(C)=\min\big\{{\rm wt}({\bf c})\,\big|\,{\bf c}\in C \backslash\{{\bf 0}\}\big\}$.
For the well-known Gray map $\phi: \mathbb{Z}_4\to\mathbb{Z}_2^2$,
$\phi(0)=(0,0)$, $\phi(1)=(0,1)$, $\phi(2)=(1,1)$, and $\phi(3)=(1,0)$,
there is a natural extension:
\begin{equation}
\begin{array}{cccl}
\Phi:& \mathbb{Z}_2^{\alpha}\times\mathbb{Z}_4^{\beta}
  &\to&\quad\mathbb{Z}_2^{n}, \\
 & (a, a') &\mapsto& \Phi(a, a')=\big(a_0,\cdots,a_{\alpha-1},
  \phi(a'_0), \cdots, \phi(a'_{\beta-1})\big);
\end{array}
\end{equation}
where $n=\alpha+2 \beta$,
$a=(a_0,\cdots,a_{\alpha-1})\in\mathbb{Z}_2^{\alpha}$ and
$a'=(a'_0,\cdots,a'_{\beta-1})\in\mathbb{Z}_4^{\beta}$.
The map $\Phi$ is an isometry which transforms
the weight in Eqn \eqref{weight}
to the Hamming weigh in $\mathbb{Z}_2^{n}$.
We call the fraction
\begin{equation}\label{def relat dis}
 \Delta(C)=\frac{{\rm wt}(C)}{n}
\end{equation}
the {\em relative minimum distance of $C$}.


Since $C$ is a subgroup of $\mathbb{Z}_2^{\alpha}\times\mathbb{Z}_4^{\beta}$,
it is also isomorphic to an abelian structure like
$\mathbb{Z}_2^{\gamma}\times\mathbb{Z}_4^{\delta}$.
Therefore, $C$ is of type $2^{\gamma}4^{\delta}$ as a group,
the number of elements of $C$ is $|C|=2^{\gamma+2\delta}$,
and the number of the elements of order $2$ in $C$
equals to $2^{\gamma+\delta}$.
We say that the fraction
\begin{equation}\label{def rate}
 R(C)=\frac{\gamma+2\delta}{n}
\end{equation}
is the {\em rate} of the code $C$.

\smallskip
There is an additive group homomorphism
from $\mathbb{Z}_2$ into $\mathbb{Z}_4$:
\begin{equation}\label{2-map}
 \mathbb{Z}_2\to\mathbb{Z}_4,~~ a\mapsto 2a,
\end{equation}
and the image of the homomorphism is
$2\mathbb{Z}_4=\{0,2\}\subset \mathbb{Z}_4$.
With the help of this homomorphism,
the standard inner product in $\mathbb{Z}_2^{\alpha}\times\mathbb{Z}_4^{\beta}$
is defined in \cite{BFPRV} as follows:
for any
$\mathbf{a}=(a_{0},\cdots, a_{\alpha-1}, a'_{0},\cdots, a'_{\beta-1})$
and
$\mathbf{b}=(b_{0},\cdots, b_{\alpha-1}, b'_{0},\cdots, b'_{\beta-1})$,
$$\mathbf{a}\cdot\mathbf{b}=2\left(\sum_{i=0}^{\alpha-1}a_i b_i\right)
 +\sum_{i=0}^{\beta-1}a'_i b'_i~\in~ \mathbb{Z}_{4}.$$
Let $C$ be any $\mathbb{Z}_2\mathbb{Z}_4$-additive code.
The {\em dual code} of $C$ is defined in the standard way by
$$C^{\bot}=\{\mathbf{a}\in\mathbb{Z}_2^{\alpha}\times\mathbb{Z}_4^{\beta}
  ~\mid~\mathbf{c}\cdot\mathbf{a}=0, ~\text{for all}~\mathbf{c}\in C\}.$$

Next, we introduce the definition of
$\mathbb{Z}_2\mathbb{Z}_4$-additive cyclic codes
which is a natural extension of the definition of
classical cyclic codes.

\begin{definition}\label{acyclic}\rm\cite[Definition 3]{ASA}~
A subset $C$ of $\mathbb{Z}_2^{\alpha}\times\mathbb{Z}_4^{\beta}$
is called a {\em $\mathbb{Z}_2\mathbb{Z}_4$-additive cyclic code} if
\begin{itemize}
\item[(i)] $C$ is an additive code;
\item[(ii)] For any codeword
${\bf c}=(c_0,c_{1},\cdots,c_{\alpha-1},c'_{0},c'_{1},\cdots,c'_{\beta-1})\in C$,
its cyclic shift
$\xi({\bf c})=(c_{\alpha-1},c_{0},\cdots,c_{\alpha-2},
  c'_{\beta-1},c'_{0},\cdots,c'_{\beta-2})$
is still in $C$.
\end{itemize}
\end{definition}

Note that in the definition the length $\alpha$ of the $\mathbb{Z}_2$-cycle
and the length $\beta$ of the $\mathbb{Z}_4$-cycle may be different.

Denote $R_{\alpha,\beta}=R_{\alpha}\times R'_{\beta}$,
where $R_{\alpha}=\mathbb{Z}_2[X]/\langle X^{\alpha}-1\rangle$
is the residue ring of the polynomial ring $\mathbb{Z}_2[X]$
modulo the ideal $\langle X^{\alpha}-1\rangle$ generated by
$X^{\alpha}-1$, and
$R'_{\beta}=\mathbb{Z}_4[X]/\langle X^{\beta}-1\rangle$.
Any element
${\bf a}=(a_{0},a_1,\cdots, a_{\alpha-1}, a'_{0},a'_1,\cdots, a'_{\beta-1})
 \in\mathbb{Z}_2^{\alpha}\times\mathbb{Z}_4^{\beta}$
can be identified with an element
${\bf a}(X)=(a(X), a'(X))\in R_{\alpha,\beta}$, where
$$\begin{array}{l}
a(X)= a_{0}+a_{1}X+\cdots+a_{\alpha-1}X^{\alpha-1}~\in~ R_\alpha,\\[3pt]
a'(X)= a'_{0}+ a'_{1}X+\cdots+a'_{\beta-1}X^{\beta-1}~\in~ R'_\beta.
\end{array}$$
This identification gives a one-to-one correspondence
between $\mathbb{Z}_2^{\alpha}\times\mathbb{Z}_4^{\beta}$
and $R_{\alpha,\beta}$.

For $f(X)\in \mathbb{Z}_4[X]$ and $(a(X), a'(X))\in R_{\alpha,\beta}$,
we define
$$f(X)\big(a(X), a'(X)\big)=\big(
\overline{f}(X)a(X),~ f(X)a'(X)\big)~\in~ R_{\alpha,\beta};$$
where $\overline{f}(X)$ is the reduction
$\overline{f}(X)= f(X)\pmod 2$, and
$\overline{f}(X)a(X)\in R_\alpha=\mathbb{Z}_2[X]/\langle X^{\alpha}-1\rangle$,
i.e., we should take the residue of $\overline{f}(X)a(X)$ mod $X^{\alpha}-1$;
the same is true of
$f(X)a'(X)\in R'_\beta=\mathbb{Z}_4[X]/\langle X^{\beta}-1\rangle$.
In a word, $R_{\alpha,\beta}$ is a $\mathbb{Z}_4[X]$-module.
It is shown in \cite{ASA} that
$\mathbb{Z}_2\mathbb{Z}_4$-additive cyclic codes in
$\mathbb{Z}_2^{\alpha}\times\mathbb{Z}_4^{\beta}$
can be identified with $\mathbb{Z}_4[X]$-submodules of
$R_{\alpha,\beta}=\mathbb{Z}_2[X]/\langle X^{\alpha}-1\rangle
\times\mathbb{Z}_4[X]/\langle X^{\beta}-1\rangle$.

As is common in the discussion of usual cyclic codes, we write
codewords of a $\mathbb{Z}_2\mathbb{Z}_4$-additive cyclic code $C$
as vectors or as polynomials interchangeably.
In either case, we use the same notation $C$ to denote the set of all codewords.
We follow this convention in the rest of the paper.

\section{A class of $\mathbb{Z}_2\mathbb{Z}_4-$additive cyclic codes}

From now on,
we always assume that $m$ is an odd positive integer.
For fundamentals on finite rings and coding theory,
please refer to \cite{HP, Mc}.

Following Section 2, we denote
$$R_m=\mathbb{Z}_2[X]/\langle X^{m}-1\rangle, \qquad
  R'_{m}=\mathbb{Z}_4[X]/\langle X^{m}-1\rangle, \qquad
  2\mathbb{Z}_4=\{0,2\}\subset\mathbb{Z}_4.
$$
We have seen that a subgroup $C$ of $\mathbb{Z}_2^m\times\mathbb{Z}_4^m$
is a $\mathbb{Z}_2\mathbb{Z}_4$-additive cyclic code
if and only if $C$ is a $\mathbb{Z}_4[X]$-submodule of
$R_m\times R'_m$.
These codes are a special class
of $\mathbb{Z}_2\mathbb{Z}_4$-additive cyclic codes.
We call such codes by $\mathbb{Z}_2\mathbb{Z}_4$-additive
cyclic codes with {\em cycle length $m$}.

Now we construct a class of $\mathbb{Z}_4[X]$-submodules of $R_m\times R'_m$.
Let
$$
2R'_{m}=\big\{a'(X)=\sum_{i=0}^{m-1}a'_iX^i\in R'_{m}~:~ 2\,|\,a'_i\big\},
$$
which is an ideal of $R'_m$, i.e.,
a $\mathbb{Z}_4[X]$-submodule of $R'_m$.
From the group homomorphism Eqn \eqref{2-map},
we have a $\mathbb{Z}_4[X]$-module isomorphism from $R_m$ onto $2R'_m$
as follows:
\begin{equation}\label{2R'}
 R_m \longrightarrow 2R'_m,~~
  \sum_{i=0}^{m-1}a_iX^i\longmapsto \sum_{i=0}^{m-1}2a_iX^i.
\end{equation}
Both $R_m$ and $2R'_m$ are in fact $\mathbb{Z}_2[X]$-modules
(and moreover, they are $R_m$-modules),
the above is a $\mathbb{Z}_2[X]$-isomorphism
(moreover, an $R_m$-isomorphism).

Thus, $R_m\times 2R'_{m}$ is a $\mathbb{Z}_4[X]$-submodule
of $R_m\times R'_{m}$. Hence any $\mathbb{Z}_2[X]$-submodule
of $R_m\times 2R'_{m}$,
i.e., any $R_m$-submodule of $R_m\times 2R'_{m}$,
is a $\mathbb{Z}_4[X]$-submodule of $R_m\times R'_{m}$.

Each element of $R_m\times 2R'_{m}$
is uniquely represented as
$$\big(a(X),2b(X)\big)\quad {\rm with}\quad
a(X)=\sum_{j=0}^{m-1}a_{j}X^{j},~
b(X)=\sum_{j=0}^{m-1}b_{j}X^{j}\in R_m.
$$
Any $R_m$-submodule of $R_m\times 2R'_{m}$
is generated by at most two elements. There are submodules of it
which cannot be generated by one element.
For example, $R_m\times 2R'_{m}$ as an $R_m$-submodule can not be generated
by one element, because: any $R_m$-module generated by one element
is of dimension over $\mathbb{Z}_2$ at most $m$.

For any given $\big(a(X),b(X)\big)\in R_m\times R_{m}$, let
\begin{equation}\label{C_a,b}
C_{a,b}=\big\{\big(f(X)a(X),2f(X)b(X)\big)\in R_m\times 2R'_{m}
 \,\big|\,f(X)\in R_m\big\},
\end{equation}
which is obviously an $R_m$-submodule of $R_m\times 2R'_{m}$,
hence is a $\mathbb{Z}_4[X]$-submodule of $R_m\times R'_{m}$;
in other words, $C_{a,b}$ is a $\mathbb{Z}_2\mathbb{Z}_4$-additive
cyclic code in $R_m\times R'_{m}$
generated by one element $\big(a(X),b(X)\big)$.
Since $C_{a,b}$ is an $R_m$-module,
it is a $\mathbb{Z}_2$-vector space,
and its dimension is denoted by $\dim C_{a,b}$.
We will exhibit good algebraic structures and
good asymptotic properties of the
$\mathbb{Z}_2\mathbb{Z}_4$-additive cyclic codes
with cycle length $m$ defined above.

We show a way to construct a generator matrix
of the code $C_{a,b}$ in Eqn~\eqref{C_a,b}.
Recall that the element $\big(a(X),2b(X)\big)\in R_m\times 2R'_{m}$
is always identified with the word
$$(a_{0},a_{1},\cdots,a_{m-2},a_{m-1},~2b_{0},2b_{1},\cdots,2b_{m-2},2b_{m-1})
 \in \mathbb{Z}_2^m\times \mathbb{Z}_4^m. $$
We write
$$\begin{array}{l}
a(X)=a_0+a_1X+\cdots+ a_{m-1}X^{m-1},\\[5pt]
b(X)=b_0+b_1X+\cdots+b_{m-1}X^{m-1}.
\end{array}$$
For $a(X)$ we have an $m$-dimensional vector $(a_0,a_1,\cdots,a_{m-2},a_{m-1})$,
from which a circulant $m\times m$ matrix is constructed as follows:
$$
A=\begin{pmatrix}a_0 & a_1 & \cdots & a_{m-2} & a_{m-1} \\
 a_{m-1} & a_0 &\cdots &a_{m-3}& a_{m-2} \\
 \cdots & \cdots & \cdots & \cdots & \cdots\\
 a_{2} & a_{3} & \cdots & a_0& a_{1}\\
 a_{1}  & a_{2} & \cdots & a_{m-1} & a_0
\end{pmatrix}_{m\times m}.
$$
Similarly, from $b(X)$ we have a circulant $m\times m$ matrix as follows:
$$
B=\begin{pmatrix}b_0 & b_1 & \cdots & b_{m-2} & b_{m-1} \\
 b_{m-1} & b_0 &\cdots &b_{m-3}& b_{m-2} \\
 \cdots & \cdots & \cdots & \cdots & \cdots\\
 b_{2} & b_{3} & \cdots & b_0& b_{1}\\
 b_{1}  & b_{2} & \cdots & b_{m-1} & b_0
\end{pmatrix}_{m\times m}.
$$
Then we can construct an $m\times 2m$ matrix:
\begin{equation}\label{A A'}
 \widehat A=\left(\begin{array}{cc}
  A& 2B
 \end{array}\right)_{m\times 2m}.
\end{equation}
And it is easy to see that
$$
C_{a,b}=
\big\{(y_{0},y_1\cdots,y_{m-1})\widehat A\in
 \mathbb{Z}_2^{m}\times \mathbb{Z}_4^m
\,\big|\,(y_{0},y_1\cdots,y_{m-1})\in \mathbb{Z}_2^{m}\big\}.
$$
However, $\widehat A$ is not a generator matrix of $C_{a,b}$ in general,
since it is not of rank~$m$ in general.
To get a generator matrix from $\widehat A$,
we state a theorem as follows,
where $\gcd(\cdot,\cdot)$ denotes the greatest common divisor
and $\deg f(X)$ denotes the degree of the polynomial.

\begin{theorem}\label{thm2}
Given $(a(X),b(X))\in R_m\times R_{m}$ and $C_{a,b}$ as in Eqn \eqref{C_a,b}.
Let
\begin{equation}\label{g_a,a'}
 g_{a,b}(X)=\gcd\big(a(X),b(X),X^m-1\big)\quad \mbox{and}\quad
 h_{a,b}(X)=\frac{X^{m}-1}{g_{a,b}(X)}.
\end{equation}
Let $\big\langle g_{a,b}(X)\big\rangle_{R_m}$ be
the ideal of $R_m$ generated by $g_{a,b}(X)$.
Then $\dim C_{a,b}=\deg h_{a,b}(X)$, and
there is an $R_m$-module isomorphism
$\big\langle g_{a,b}(X)\big\rangle_{R_m}\cong C_{a,b}$
which maps $c(X)\in\big\langle g_{a,b}(X)\big\rangle_{R_m}$ to
$\big(c(X)a(X),\,2c(X)b(X)\big)\in C_{a,b}$.
\end{theorem}

\pf The following map is clearly an $R_{m}$-module homomorphism:
$$\gamma_{a,b}:~
 R_m\longrightarrow R_m\times 2R'_{m},\quad
 f(X)\longmapsto \big(f(X)a(X),\;2f(X)b(X)\big).
$$
Obviously, the image ${\rm im}(\gamma_{a,b})=C_{a,b}$.
Next we compute the kernel ${\rm ker}(\gamma_{a,b})$.
For $f(X)\in R_m$, $f(X)\in{\rm ker}(\gamma_{a,b})$
if and only if in $\mathbb{Z}_2[X]$ we have
\begin{equation}\label{system}
\left\{\begin{matrix}f(X)a(X)\equiv 0\pmod{X^{m}-1},\\
 f(X)b(X)\equiv 0\pmod{X^{m}-1}.\end{matrix}\right.
\end{equation}
The two equations are combined into one equation in $\mathbb{Z}_2[X]$:
$$f(X)\gcd\big(a(X),b(X)\big)\equiv 0~({\rm mod}~X^m-1),$$
from which we see that the system \eqref{system} holds if and only if
$$\textstyle
f(X)\equiv 0\quad \big({\rm mod}~\frac{X^{m}-1}{\gcd(a(X),b(X),X^m-1)}\big).
$$
So we get that
${\rm ker}(\gamma_{a,b})=\big\langle h_{a,b}(X)\big\rangle_{R_m}$.
Since $m$ is odd, $R_m$ is semisimple, i.e.,
$X^{m}-1$ has no repeated (multiple) roots in any extension of $\mathbb{Z}_2$.
Then we obtain a direct sum decomposition of $R_m$ as follows:
$$R_m=\big\langle h_{a,b}(X)\big\rangle_{R_m}\oplus
 \big\langle g_{a,b}(X)\big\rangle_{R_m}.
$$
Then the above $R_m$-homomorphism $\gamma_{a,b}$
induces an $R_m$-isomorphism:
$$\overline\gamma_{a,b}:~
 \big\langle g_{a,b}(X)\big\rangle_{R_m}
  \mathop{\longrightarrow}^{\cong} C_{a,b},\quad
 c(X)\longmapsto \big(c(X)a(X),\;2c(X)b(X)\big).
$$
In particular,
$$\dim C_{a,b}=\dim\big\langle g_{a,b}(X)\big\rangle_{R_m}
 =m-\deg g_{a,b}(X)=\deg h_{a,b}(X).\eqno\qed
$$

\begin{corollary}\label{cor3.2}
Let $C_{a,b}$, $h_{a,b}(X)$ be as in Theorem \ref{thm2} and
$r=\deg h_{a,b}(X)$, let $\widehat A$ be as in Eqn \eqref{A A'}.
Then the matrix consisting of the first $r$ rows of $\widehat A$
is a generator matrix of $C_{a,b}$.
\end{corollary}

\pf From the above theorem, we see that
the rank of $\widehat A$ is equal to $r$.
And by \cite[Theorem 3.6]{FL17}, the first $r$ rows of $\widehat A$
are linearly independent.
\qed

\begin{example}\label{exa1}\rm
 Take $m=3$, $a(X)=(X+1)^2=X^2+1$
and $b(X)=X+1$. By Theorem \ref{thm2},
$$\begin{array}{ll}
g_{a,b}(X)=\gcd\big(a(X),b(X),X^3-1\big)=X+1;\\[5pt]
 h_{a,b}(X)=(X^{3}-1)/g_{a,b}(X)=X^2+X+1.
\end{array}$$
So $\dim C_{a,b}=2$. Using the notations in Eqn \eqref{A A'}, we have
$$
 A=\begin{pmatrix}1&0&1\\ 1&1&0\\ 0&1&1 \end{pmatrix},
 \qquad B=\begin{pmatrix}1&1&0\\ 0&1&1\\1&0&1 \end{pmatrix};
$$
$$
\widehat A=\begin{pmatrix}1&0&1&2&2&0\\ 1&1&0&0&2&2\\ 0&1&1&2&0&2  \end{pmatrix}.
$$
Thus the first two rows of $\widehat A$ are linearly independent and
$$
 G=\begin{pmatrix} 1&0&1&2&2&0\\ 1&1&0&0&2&2\end{pmatrix}
$$
is a generator matrix of $C_{a,b}$. The $4$ codewords of $C_{a,b}$ are
as follows:
$$\begin{array}{cccc}
000000, & 110022,&101220, &011202.
\end{array}$$
\end{example}

\begin{example}\label{exa2}\rm
 Take $m=3$, $a(X)=X^2+X+1$
and $b(X)=X+1$. By Theorem \ref{thm2},
$$\begin{array}{ll}
g_{a,b}(X)=\gcd\big(a(X),b(X),X^3-1\big)
  =1;\\[5pt]
h_{a,b}(X)=(X^{3}-1)/g_{a,b}(X)=X^3-1.
\end{array}$$
So $\dim C_{a,b}=3$. And by the notations in Eqn \eqref{A A'},
$$
 A=\begin{pmatrix}1&1&1\\ 1&1&1\\ 1&1&1 \end{pmatrix},
 \qquad B=\begin{pmatrix}1&1&0\\ 0&1&1\\ 1&0&1\end{pmatrix};
$$
$$
\widehat A=\begin{pmatrix}1&1&1&2&2&0\\ 1&1&1&0&2&2\\ 1&1&1&2&0&2  \end{pmatrix}.
$$
Thus the rows of $\widehat A$ are linearly independent and
$$
 G=\left(\begin{array}{cccccc}
1&1&1&2&2&0\\ 1&1&1&0&2&2\\ 1&1&1&2&0&2
\end{array}\right)
$$
is a generator matrix of $C_{a,b}$. The $8$ codewords of $C_{a,b}$ are
as follows:
$$\begin{array}{cccccccc}
000000, & 111202, &111022, &000220,&111220,&000022, &000202, &111000.
\end{array}$$
\end{example}

Recall that $m$ is odd, for $X^m-1=(X-1)(X^{m-1}+\cdots+X+1)$, we get
\begin{equation}\label{a decomp Rm}
R_m=\mathbb{Z}_2[X]/\langle X^m-1\rangle
=\langle X^{m-1}+\cdots+X+1\rangle_{R_m}\oplus \langle X-1 \rangle_{R_m},
\end{equation}
where $\langle X^{m-1}+\cdots+X+1\rangle_{R_m}$ is the trivial code
of dimension $1$. Thus, in the following we
consider the following ideal of $R_m$ generated by $X-1$:
\begin{equation}\label{J}
  J_m=\langle X-1 \rangle_{R_m},
\end{equation}
which is called the {\em augmentation ideal} of $R_{m}$ in
the representation theory of finite groups.
Similarly, in $R'_m=\mathbb{Z}_4[X]/\langle X^m-1\rangle$
we denote $J'_m=\langle X-1\rangle_{R'_m}$.
It is easy to see that
$2J'_m=\{2f(X)\mid f(X)\in J'_m\}=\langle 2(X-1)\rangle_{R'_m}$,
and the $R_m$-isomorphism in Eqn \eqref{2R'}
induces an $R_m$-module isomorphism:
\begin{equation}\label{JJ'}
 J_m\times J_m \mathop{\longrightarrow}^{\cong} J_m\times 2J'_m,\quad
   (a(X),b(X))\longmapsto (a(X),2b(X)).
\end{equation}

For the code $C_{a,b}$ constructed in Eqn~\eqref{C_a,b}
where $a(X)$, $b(X)$ chosen from $R_m$,
in the following we will consider
only the $a(X)$, $b(X)$ chosen from $J_{m}$.
In that case when $a(X), b(X)\in J_m$, by Theorem \ref{thm2},
the $\mathbb{Z}_2\mathbb{Z}_4$-additive cyclic code $C_{a,b}$
with cycle length $m$
can be reformulated within $J_{m}$ as follows:
\begin{equation}\label{JC_a,b}
C_{a,b}=\big\{(c(X)a(X),2c(X)b(X))\in R_m\times 2R'_{m}
 \;\big|\;c(X)\in J_{m}\big\}\subseteq J_m\times 2J'_m.
\end{equation}

The class of the $\mathbb{Z}_2\mathbb{Z}_4$-additive cyclic code 
 $C_{a,b}$ (of cycle length $m$) for $a(X),b(X)\in J_m$ 
 will play a key role in the next section. 

\begin{lemma}\label{dC_a,b}
Let notation be as above,
let $\big(a(X),b(X)\big)\in J_{m}\times J_{m}$. Then
$\dim C_{a,b}\le m-1$; and $\dim C_{a,b}<m-1$
if and only if there is an irreducible factor
$p(X)$ of $X^{m-1}+\cdots+X+1$ in $\mathbb{Z}_2[X]$
such that $p(X)\,|\,a(X)$ and $p(X)\,|\,b(X)$.
\end{lemma}

\pf By the definition of $J_{m}$ in Eqn~\eqref{J} we have
$$(X-1)\,\big|\;g_{a,b}(X)=\gcd\big(a(X),b(X),X^m-1\big);$$
where $g_{a,b}(X)$ is defined in Eqn \eqref{g_a,a'}. So
$$
\big\langle g_{a,b}(X)\big\rangle_{R_m}\subseteq
\big\langle X-1\big\rangle_{R_m}=J_{m}.
$$
By Theorem \ref{thm2},
$\dim C_{a,b}=\dim\big\langle g_{a,b}(X)\big\rangle_{R_m}\le\dim J_m=m-1$.
And,
$\dim C_{a,b}=m-1$ if and only if
$$g_{a,b}(X)=\gcd\big(a(X),b(X),X^m-1\big)=X-1,$$
if and only if there is no irreducible factor
$p(X)$ of $\frac{X^m-1}{X-1}$
such that $p(X)\,|\,a(X)$ and $p(X)\,|\,b(X)$.
\qed

\begin{lemma}\label{number I}
Let $\ell_m$ be the minimal degree of irreducible factors
in $\mathbb{Z}_2[X]$ of $\frac{X^m-1}{X-1}$,
and $d$ be an integer with $\ell_m\le d<m$.
Then any non-zero ideal of $R_m$ contained in $J_{m}$
is of dimension at least $\ell_m$, and the number of the ideals
of dimension $d$ which are contained in $J_{m}$
is at most $m^{\frac{d}{\ell_m}}$.
\end{lemma}

\pf By the decomposition in Eqn \eqref{a decomp Rm}
and the definition of $J_m$ in Eqn \eqref{J},
each irreducible ideal contained in $J_{m}$ is corresponding to exactly
one irreducible divisor of $\frac{X^m-1}{X-1}$ such that the dimension
of the ideal is equal to the degree of the corresponding divisor. Thus,
the minimal dimension of the ideals contained in $J_{m}$ is equal to $\ell_m$.
And, any $d$-dimensional ideal contained in $J_{m}$
is a sum of at most $d/\ell_m$ irreducible ideals.
So the number of the $d$-dimensional ideals contained in $J_{m}$
is at most the partial sum of binomial coefficients
$\sum_{i=1}^{\lfloor d/\ell_m\rfloor}{h\choose i}$,
where $h$ is the number of the irreducible ideals contained in $J_{m}$
 and $\lfloor d/\ell_m\rfloor$ denotes
the largest integer which is not larger than $d/\ell_m$.
It is easy to check that
\begin{eqnarray*}
\sum_{i=1}^{\lfloor d/\ell_m\rfloor}{h\choose i}&\leq& \sum_{i=1}^{\lfloor d/\ell_m\rfloor}{m\choose i}\\
&\leq&\frac{d}{\ell_m}\cdot\frac{m(m-1)\cdots(m-\frac{d}{\ell_m}+1)}{1\cdot2\cdots\frac{d}{\ell_m}}\\
&\leq&m^{\frac{d}{~\ell_m}}~.
\end{eqnarray*}
\qed

\section{Random $\mathbb{Z}_2\mathbb{Z}_4-$additive cyclic codes}

Keep the notations in Section 2.

For a real number $x$ with $0\le x\le 1$, let
 $H(x)=-x\log_2 x-(1-x)\log_2(1-x)$
with the convention that $0\log_q0=0$.
The function $H(x)$ is called the (binary) {\em entropy}.
Note that $H(x)$ is a strictly increasing concave function
in the interval $[0,1/2]$ with $H(0)=0$ and $H(1/2)=1$,
see \cite[\S 2.10.6]{HP}.
In the following, we assume that $\delta$ is a real number such that
\begin{equation}\label{delta}
0<\delta< 1/3 \qquad\mbox{and}\qquad H(3\delta/2)< 1/2.
\end{equation}
Note that, if $x\in[0,1/2]$ such that $H(x)=1/2$, then $0.11<x<0.1101$.
So the assumption (4.1) is about to say that  
$0<\delta\le 0.0733$.

As in Section 3, we denote
$$ R_m=\mathbb{Z}_2[X]/\langle X^m-1\rangle,\qquad
 R'_m=\mathbb{Z}_4[X]/\langle X^m-1\rangle,
$$
and (see Eqn \eqref{J} and Eqn \eqref{JJ'})
$$J_m=\langle X-1\rangle_{R_m}, \qquad
2J'_m=\langle 2(X-1)\rangle_{R'_m}=\{2b(X)\mid b(X)\in J_m\}.$$
In this section we view the set $J_{m}\times 2J'_{m}$
as a probability space, whose samples are afforded with equal probability.
For each sample $\big(a(X),2b(X)\big)\in J_{m}\times 2J'_{m}$
(with $b(X)\in J_m$), by Eqn \eqref{JC_a,b}
we have a $\mathbb{Z}_2\mathbb{Z}_4-$additive cyclic code
with cycle length $m$:
\begin{equation}\label{JJC_a,b}
C_{a,b}=\big\{(c(X)a(X),2c(X)b(X))\in R_m\times 2R'_{m}
 \,\big|\,c(X)\in J_{m}\big\}.
\end{equation}
Then $C_{a,b}$ is a random code over
the probability space $J_{m}\times 2J'_{m}$,
hence the relative minimum distance $\Delta(C_{a,b})$ of $C_{a,b}$
 (see Eqn \eqref{def relat dis}) and the dimension $\dim(C_{a,b})$ of $C_{a,b}$
are random variables over the probability space.
We are interested in the asymptotic properties
of the probabilities ${\Pr\big(\Delta(C_{a,b})>\delta\big)}$
and ${\Pr\big(\dim C_{a,b}=m-1\big)}$.
As a preparation, we introduce a type of $0$-$1$-variables.

Recall that a random variable  is called a {\em Bernoulli variable} 
if it takes the value $1$ with probability $p$ 
and the value $0$ with probability $1-p$. 
Given any $c(X)\in J_{m}$. We define a Bernoulli variable $Y_c$
over the probability space $J_{m}\times 2J'_{m}$ as follows:
for samples $\big(a(X),2b(X)\big)\in J_{m}\times 2J'_{m}$,
\begin{equation}\label{Y_c}
Y_c=\begin{cases} 1, & 1\le{\rm wt}\big(c(X)a(X),2c(X)b(X)\big)\le 3m\delta;\\
 0, & {\rm otherwise}.\end{cases} \quad
\end{equation}
Since $c(X)\in J_{m}$, the set
$\{c(X)a(X)\in R_m\,|\,a(X)\in J_{m}\}$ is the ideal of $R_m$
generated by $c(X)$, we denote it by $I_c$ and denote its dimension by $d_c$,
i.e.,
\begin{equation}\label{I_c}
 I_c=\big\langle c(X)\big\rangle_{R_m}\subseteq J_{m}, \quad d_c=\dim I_c\,.
\end{equation}

By $|S|$ we denote the cardinality of any set $S$.

\begin{lemma}\label{IcIc}
Let notation be as in Eqns \eqref{delta}--\eqref{I_c}.
For $I_c\times I_c\subseteq R_m\times R_m$, let
$$(I_c\times I_c)^{\le 3m\delta}=
  \big\{\big(f(X),g(X)\big)\in I_c\times I_c\,\big|\;
  {\rm wt}_H\big(f(X),g(X)\big)\le 3m\delta \big\}.$$
Then
$$ \big|(I_c\times I_c)^{\le 3m\delta}\big|\le 2^{2d_c H(3\delta/2)}. $$
\end{lemma}

\pf Note that the length of the code
$I_c\times I_c$ is $2m$, i.e., $\dim(R_m\times R_m)=2m$.
And the dimension $\dim(I_c\times I_c)=2d_c$.
The fraction of $3m\delta$ over the length is
$(3m\delta)/(2m)=3\delta/2$, and by the assumption Eqn \eqref{delta},
$0<3\delta/2<1/2$. So
the inequality follows immediately from
\cite[Remark~3.2 and Corollary~3.5]{FL}.
\qed

With the notation as in Eqns \eqref{delta}--\eqref{I_c},
we show an estimation of the expectation ${\rm E}(Y_c)$
of the random variable $Y_c$ for later quotation.

\begin{lemma}\label{EY_c}~
${\rm E}(Y_c)\le 2^{-2d_c+2d_c H(3\delta/2)}$.
\end{lemma}

\pf
In $2J'_{m}\subset R'_m$, we have an ideal
$$
 I'_c=\big\langle 2c(X)\big\rangle_{R'_{m}}
 =\{2c(X)b(X)\in R'_{m}\,|\,b(X)\in J_{m}\} \subseteq 2J'_{m}.
$$
By Eqn \eqref{JJ'}, we see that:
\begin{equation}\label{Ic=I'c}
I'_c\cong I_c, \qquad\mbox{and}\qquad
 \dim I'_c=\dim I_c=d_c\,.
\end{equation}
For $I_c\times I'_c\subseteq R_m\times R'_m$, we denote
$$(I_c\times I'_c)^{\le 3m\delta}=
  \big\{\big(f(X),2g(X)\big)\in I_c\times I'_c\,\big|\;
  {\rm wt}\big(f(X),2g(X)\big)\le 3m\delta \big\}.$$
Since $Y_c$ is a $0$-$1$-variable, the expectation of $Y_c$
is just the probability that $Y_c=1$, So
\begin{equation}\label{EYc}
 {\rm E}(Y_c)=\Pr(Y_c=1)=
 \frac{\big|(I_c\times I'_c)^{\le 3m\delta}\big|-1}
 {|I_c\times I'_c|}.
\end{equation}
For $f(X),g(X)\in R_m$, we have seen that
$${\rm wt}_L(2g(X))=2\,{\rm wt}_H(g(X)),$$
where the weight in the left hand side is in $R'_m$,
while the weight in the right hand side is in $R_m$; so
\begin{equation*}
{\rm wt}(f(X),2g(X))\ge {\rm wt}_H(f(X),g(X)).
\end{equation*}
Thus, with notation in Lemma \ref{IcIc}, we get
\begin{equation}\label{ww}
\big|(I_c\times I'_c)^{\le 3m\delta}\big|
\le \big|(I_c\times I_c)^{\le 3m\delta}\big|.
\end{equation}
By Eqn \eqref{EYc}, Eqn \eqref{ww}, Eqn \eqref{Ic=I'c} and Lemma \ref{IcIc},
$$
 {\rm E}(Y_c)\le
 \frac{\big|(I_c\times I_c)^{\le 3m\delta}\big|}
 {|I_c\times I_c|}\le \frac{2^{2d_cH(3\delta/2)}}{2^{2d_c}}
 =2^{-2d_c+2d_c H(3\delta/2)}.
 \eqno\qed
$$

Now we can present an estimation of the probability
that the relative minimum distance $\Delta(C_{a,b})$ is at most $\delta$.

\begin{lemma}\label{EX}
Let $\delta$ be as in Eqn \eqref{delta} and $C_{a,b}$ be as in
Eqn \eqref{JJC_a,b}. Let $\ell_m$ be the minimal degree of
the irreducible divisors of $\frac{X^m-1}{X-1}$ as in Lemma \ref{number I}.
Then
$$ \Pr\big(\Delta(C_{a,b})\le\delta\big)\le
\sum_{j=\ell_m}^{m-1}
 2^{-2j\big(\frac{1}{2}-H(\frac{3\delta}{2})-\frac{\log_2 m}{\ell_m}\big)}.
$$
\end{lemma}

\pf
Let $Y_c$ for $c(X)\in J_{m}$ be the $0$-$1$-variable
defined in Eqn \eqref{Y_c}. Let
$$Y=\sum_{c(X)\in J_{m}} Y_c\,.$$
Then $Y$ is a non-negative integer random variable
over the probability space $J_{m}\times 2J'_{m}$.
By Eqn \eqref{JJC_a,b} and Eqn \eqref{Y_c}, $Y$ stands for
the number of $c(X)\in J_{m}$ such that the codeword
$\big(c(X)a(X), 2c(X)b(X)\big)$ of $C_{a,b}$ is non-zero with weight
at most $3m\delta$. By Eqn \eqref{def relat dis},
$\Delta(C_{a,b})={\rm wt}(C_{a,b})/3m$. We get
$$
 \Pr\big(\Delta(C_{a,b})\le\delta\big)=\Pr\big(Y>0\big).
$$
By a Markov's inequality \cite[Theorem 3.1]{MU},
$\Pr\big(Y>0\big)\le {\rm E}(Y)$. So we can prove the lemma by
estimating the expectation ${\rm E}(Y)$.

By the linearity of the expectation,
${\rm E}(Y)=\sum_{c(X)\in J_{m}}{\rm E}(Y_c)$.
For any ideal $I$ of $J_{m}$ (we denote it by $I\le J_{m}$), let
$I^*=\big\{c(X)\in I\,\big|\,I_c=I\big\}$, where
$I_c$ is defined in Eqn \eqref{I_c}. That is,
$$
I^*=\big\{c(X)\in I\,\big|\,d_c=\dim I\big\},
$$
where $d_c=\dim I_c$, see Eqn \eqref{I_c}.
It is easy to see that
$ J_{m}=\bigcup_{I\le J_{m}}I^*$, where the subscript
``$I\le J_{m}$'' means that $I$ runs over the ideals contained in $J_{m}$.
By Lemma \ref{number I}, if $0\ne I\le J_{m}$
then $\ell_m\le\dim I\le m-1$. So
$$
{\rm E}(Y)=\sum_{I\le J_{m}}\sum_{c(X)\in I^*}{\rm E}(Y_c)
=\sum_{j=\ell_m}^{m-1}
  \sum_{\mbox{\tiny$\begin{array}{c}I\le J_{m}\\ \dim I=j\end{array}$}}
   \sum_{c(X)\in I^*}{\rm E}(Y_c).
$$
For $I\le J_{m}$ with $\dim I=j$, by Lemma \ref{EY_c}
and the fact that $|I^*|\le |I|=2^j$, we get
$$
\sum_{c(X)\in I^*}{\rm E}(Y_c)\le \sum_{c(X)\in I^*}
  2^{-2j+2j H(3\delta/2)} \le 2^{-j+2j H(3\delta/2)}.
$$
By Lemma \ref{number I} again, the number of
$I\le J_{m}$ with $\dim I=j$ is less that $m^{j/\ell_m}$.
And note that $\log_2 m\le\frac{j\log_2 m}{\ell_m}$ (as $j\ge\ell_m)$.
So
\begin{eqnarray*}
{\rm E}(Y)&\le& \sum_{j=\ell_m}^{m-1} m^{j/\ell_m} 2^{-j+2j H(3\delta/2)}\\
&=& \sum_{j=\ell_m}^{m-1}
 2^{-2j\big(\frac{1}{2}- H(\frac{3\delta}{2})-\frac{\log_2 m}{\ell_m}\big)}.
\end{eqnarray*}
The lemma is proved.
\qed

\medskip
By \cite[Lemma 2.6]{BM}, there are positive integers $m_1$, $m_2$, $\cdots$
satisfying that
\begin{equation}\label{m_i}
m_i ~\text{is odd }, \quad m_i\to\infty,\quad\lim\limits_{i\to\infty}\frac{\log_2 m_i}{\ell_{m_i}} =0,
\end{equation}
where $\ell_{m_i}$ is the minimal degree of
the irreducible divisors of $\frac{X^{m_i}-1}{X-1}$
as defined in Lemma \ref{number I}.
For each $m_i$, let
\begin{equation}\label{JJC_a,bi}
C^{(i)}_{a,b}=\big\{(c(X)a(X),2c(X)b(X))\in R_{m_i}\times 2R'_{m_i}
 \,\big|\,c(X)\in J_{m_i}\big\}
\end{equation}
be the random $\mathbb{Z}_2\mathbb{Z}_4-$additive cyclic code
with cycle length $m_i$ as defined in Eqn~\eqref{JJC_a,b}.

\begin{theorem}\label{Delta C}
Let notation be as above. Assume that $0<\delta<1/3$ and
$H(3\delta/2)<1/2$. Then
$$
 \lim\limits_{i\to\infty}\Pr\big(\Delta(C^{(i)}_{a,b})>\delta\big)=1.
$$
\end{theorem}

\pf
By the assumption (i.e. the assumption in Eqn \eqref{delta}), we have
$\frac{1}{2}-H(\frac{3\delta}{2})>0$.
By Eqn~\eqref{m_i}, there are a positive real number $\beta$
and an integer $N$ such that
$$\textstyle
 \frac{1}{2}-H(\frac{3\delta}{2})-\frac{\log_2 m_i}{\ell_{m_i}}
 \ge \beta,\qquad \forall~ i>N.
$$
By Lemma \ref{EX},
\begin{eqnarray*}
\lim\limits_{i\to\infty}\Pr\big(\Delta(C^{(i)}_{a,b})\le\delta\big)
\kern-3pt&\le&\kern-3pt\lim\limits_{i\to\infty}
\sum_{j=\ell_{m_i}}^{m_i-1}2^{-2j\beta}
\le \lim\limits_{i\to\infty}
\sum_{j=\ell_{m_i}}^{m_i-1}2^{-2\ell_{m_i}\beta}\\
&\le&\kern-3pt\lim\limits_{i\to\infty}m_i 2^{-2\ell_{m_i}\beta}
=\lim\limits_{i\to\infty}
2^{-2\ell_{m_i}\big(\beta-\frac{\log_2 m_i}{2\ell_{m_i}}\big)}.
\end{eqnarray*}
Since $\lim\limits_{i\to\infty}\frac{\log_2 m_i}{2\ell_{m_i}}=0$
(which implies that $\lim\limits_{i\to\infty}\ell_{m_i}=\infty$),
we obtain that
$\lim\limits_{i\to\infty}\Pr\big(\Delta(C^{(i)}_{a,b})\le \delta)=0$.
\qed

Next, we estimate the rate
$R(C^{(i)}_{a,b})$ of the random code $C^{(i)}_{a,b}$.

\begin{theorem}\label{dim C}
Let $m_1$, $m_2$, $\cdots$ be positive integers satisfying Eqn \eqref{m_i},
and let $C^{(i)}_{a,b}$ be as in Eqn \eqref{JJC_a,bi}. Then
$$
 \lim\limits_{i\to\infty}\Pr\big(\dim C^{(i)}_{a,b}=m_i-1\big)=1.
$$
\end{theorem}

\pf For any given $i$,
let $X^{m_i-1}+\cdots+X+1=p_1(X)\cdots p_{h_i}(X)$
be the irreducible decomposition in $\mathbb{Z}_2[X]$,
i.e., $p_1(X),\cdots p_{h_i}(X)$ are distinct irreducible polynomials
in $\mathbb{Z}_2[X]$. By Chinese Remainder Theorem,
there is a natural isomorphism:
\begin{equation}\label{J iso}
R_{m_i}\cong
\mathbb{Z}_2[X]/\langle X-1\rangle\times
 \mathbb{Z}_2[X]/\langle p_1(X)\rangle\times\cdots
  \times \mathbb{Z}_2[X]/\langle p_{h_i}(X)\rangle.
\end{equation}
We restrict it to the augmentation ideal $J_{m_i}\subset R_{m_i}$
and get a natural isomorphism:
$$
J_{m_i}=\langle X-1\rangle_{R_{m_i}}\cong
 \mathbb{Z}_2[X]/\langle p_1(X)\rangle\times\cdots
  \times \mathbb{Z}_2[X]/\langle p_{h_i}(X)\rangle
$$
which maps $a(X)\in J_{m_i}$ to
$$
\big(\mu_{m_i}^{(1)}(a(X)),\cdots,\mu_{m_i}^{(h_i)}(a(X))\big)\in
\mathbb{Z}_2[X]/\langle p_1(X)\rangle\times\cdots
  \times \mathbb{Z}_2[X]/\langle p_{h_i}(X)\rangle,
$$
where
$$
 \mu_{m_i}^{(j)}(a(X))=a(X)~({\rm mod}~{p_j(X)})~~ \in~
  \mathbb{Z}_2[X]/\langle p_j(X)\rangle, \qquad j=1,\cdots, h_i.
$$
Let $a(X),b(X)\in J_{m_i}$. By Lemma \ref{dC_a,b},
$\dim C^{(i)}_{a,b}\le m_i-1$, and $\dim C^{(i)}_{a,b}<m_i-1$
if and only if there is an index $j$ with $1\le j\le h_i$ such that
$\mu_{m_i}^{(j)}(a(X))=0=\mu_{m_i}^{(j)}(b(X))$.
In other words, $\dim C^{(i)}_{a,b}=m_i-1$
if and only if for any $j=1,\cdots,h_i$, in
$\mathbb{Z}_2[X]/\langle p_j(X)\rangle
  \times\mathbb{Z}_2[X]/\langle p_j(X)\rangle$
the following holds:
\begin{equation}\label{mu,mu}
\big(\mu_{m_i}^{(j)}(a(X)),\,\mu_{m_i}^{(j)}(b(X))\big)\ne (0,0).
\end{equation}
Let $d_j=\deg p_j(X)$. Then $\mathbb{Z}_2[X]/\langle p_j(X)\rangle$
is a finite field of cardinality $2^{d_j}$.
Note that
$\mu_{m_i}^{(j)}: J_{m_i}\to \mathbb{Z}_2[X]/\langle p_j(X)\rangle$
is surjective homomorphism, cf. the isomorphism in Eqn \eqref{J iso}.
So the probability that Eqn~\eqref{mu,mu} holds is equal to
$\frac{2^{2d_j}-1}{2^{2d_j}}=1-2^{-2d_j}$.
Obviously,  the events that Eqn~\eqref{mu,mu}
holds for $j=1,\cdots,h_i$ are randomly independent. Thus
$$
\Pr\big(\dim C^{(i)}_{a,b}=m_i-1\big)=\prod_{j=1}^{h_i}(1-2^{-2d_j}).
$$
By definition of $\ell_{m_i}$ in Eqn \eqref{m_i},
$\ell_{m_i}\le d_j$ for $j=1,\cdots,h_i$,
hence $h_i\le \frac{m_i-1}{\ell_{m_i}}\le \frac{m_i}{\ell_{m_i}}$.
Thus $\Pr\big(\dim C^{(i)}_{a,b}=m_i-1\big)\ge
\big(1-2^{-2\ell_{m_i}}\big)^{\frac{m_i}{\ell_{m_i}}}$, i.e.,
\begin{equation*}
\Pr\big(\dim C^{(i)}_{a,b}=m_i-1\big)~\ge~
 \big(1-2^{-2\ell_{m_i}}\big)^
  {2^{2\ell_{m_i}}\cdot\frac{m_i}{\ell_{m_i}2^{2\ell_{m_i}}}}.
\end{equation*}

Now we consider the asymptotic case when $i\to\infty$.
By the assumption in Eqn \eqref{m_i},
$\lim\limits_{i\to\infty}\frac{\log_2 m_i}{\ell_{m_i}}=0$
(which implies that $\lim\limits_{i\to\infty}\ell_{m_i}=\infty$).
So
$$\lim\limits_{i\to\infty}\frac{m_i}{\ell_{m_i}2^{2\ell_{m_i}}}
 =\lim\limits_{i\to\infty} 2^{-\ell_{m_i}\big(2-
  \frac{\log_2 m_i}{\ell_{m_i}}+\frac{\log_2\ell_{m_i}}{\ell_{m_i}}\big)}
 =0.
$$
Note that $\big(1-2^{-2\ell_{m_i}}\big)^{2^{2\ell_{m_i}}}>1/4$.
We get that
$$
\lim\limits_{i\to\infty}\Pr\big(\dim C^{(i)}_{a,b}=m_i-1\big)\ge
\lim\limits_{i\to\infty}(1/4)^{\frac{m_i}{\ell_{m_i}2^{2\ell_{m_i}}}}=1.
\eqno\qed
$$

\begin{theorem}\label{a g}
Let $\delta$ be a real number
such that $0<\delta< 1/3$ and $H(3\delta/2)<\frac{1}{2}$.
Then there is a sequence of $\mathbb{Z}_2\mathbb{Z}_4$-additive cyclic
codes $C_i$ with cycle length $m_i$ for $i=1,2,\cdots$
such that $m_i\to\infty$ and the following hold.

(i)~ $\lim\limits_{i\to\infty}R(C_i)=\frac{1}{3}$;

(ii)~ $\Delta(C_i)>\delta$ for all $i=1,2,\cdots$.
\end{theorem}

\pf Note that $R(C_i)=\frac{\dim C_i}{3m_i}$, see Eqn \eqref{def rate}.
From Theorem~\ref{Delta C} and Theorem~\ref{dim C}, 
there is an integer $N\ge 0$ such that for any $i>N$ we can get a 
$\mathbb{Z}_2\mathbb{Z}_4$-additive cyclic code $C_i$ with cycle length $m_i$
such that $\Delta(C_i)>\delta$ and $R(C_i)=\frac{1}{3}-\frac{1}{3m_i}$.
Removing the first $N$ codes and relabeling the remaining codes,  
we obtain the conclusions at once.
\qed

\begin{remark}\rm
For example, if we take $H(3\delta/2)=0.4999$, then $\delta\approx 0.0733$, 
and we get a sequence $C_1,C_2,\cdots$ of $\mathbb{Z}_2\mathbb{Z}_4$-additive 
cyclic codes with cycle length going to infinity such that
$R(C_i)\to 1/3$, and $\Delta(C_i)>0.0733$ for all $i=1,2,\cdots$.
\end{remark}

\begin{remark}\rm
The probabilistic techniques used in this paper 
might be applied to other families of codes,
provided one could find a class of codes in the family 
which possess suitable properties
such that the probabilistic method can work well.
\end{remark}

\section*{Acknowledgements}
The research of the authors is supported by NSFC
with grant numbers 11271005.
The authors are grateful to the reviewers for their 
careful and suggestive comments which are
helpful in creating the improved final version.

\end{document}